\documentclass[journal]{IEEEtran}
\usepackage{csquotes}
\usepackage[nolist]{acronym}
\usepackage{pifont}
\usepackage{graphicx}
\usepackage[export]{adjustbox}
\usepackage[super]{nth}
\usepackage{cite}      
\usepackage{caption}
\usepackage{subfig}
\usepackage{float}
\usepackage{tabularx}
\usepackage{multirow}
\usepackage{booktabs}
\usepackage{amsmath}
\usepackage{threeparttable}
\usepackage[dvipsnames]{xcolor}
\usepackage[hyphens,spaces,obeyspaces]{url}
\usepackage[super]{nth}
\newcolumntype{C}{>{\centering\arraybackslash}X} 
\ifCLASSINFOpdf
\else
\fi
\begin{document}
\begin{acronym}
\acro {DNN}{deep neural network}
\acro {GAN}{generative adversarial network}
\acro {GNN}{graph neural network}
\acro {DLA}{deep learning accelerator}
\acro {PVA}{programmable vision accelerator}
\acro {VIC}{video image compositor}
\acro {SSD}{single shot detector}
\acro {GPC}{graphic processing clusters}
\acro {SM}{streaming multiprocessors}
\acro {DMA}{direct memory access}
\acro {VLIW}{very long instruction word}
\acro {VPU}{vector processing unit}
\acro {TPU}{tensor processing unit}
\acro {VPI}{Vision Programming Interface}
\acro {AxoNN}{energy-aware execution of neural networks}
\acro {HaX-CoNN}{heterogeneity aware execution of concurrent deep neural networks}
\acro {D-HaX-CoNN}{dynamic heterogeneity aware execution of concurrent deep neural networks}
\acro {CP-CNN}{computational parallelization for convolutional neural network}
\acro {PCCS}{processor-centric contention-aware slowdown model}
\acro {PND}{partial network duplication}
\acro {SMT}{satisfiability modulo}
\acro {SAT}{satisfiability}
\acro {LP}{linear programming}
\acro {Jedi}{Jetson-aware embedded deep learning inference}
\acro {RNN}{recurrent neural network}
\acro {GA}{genetic algorithm}
\acro {CAN}{control area network}
\acro {H2H}{heterogeneous model to heterogeneous system mapping}
\acro {MaGNAS}{mapping-aware graph neural architecture search}
\acro {AI-SoC}{AI-System-on-Chip}
\acro {LLM}{large language models}
\acro {RL}{Reinforcement Learning}
\acro {SDRAM}{synchronous dynamic random-access memory}
\end{acronym}
\captionsetup[figure]{justification=centering}
%
% paper title
% Titles are generally capitalized except for words such as a, an, and, as,
% at, but, by, for, in, nor, of, on, or, the, to and up, which are usually
% not capitalized unless they are the first or last word of the title.
% Linebreaks \\ can be used within to get better formatting as desired.
% Do not put math or special symbols in the title.
\title{Scheduling Techniques of AI Models on Modern Heterogeneous Edge GPU - A Critical Review}
%
%
% author names and IEEE memberships
% note positions of commas and nonbreaking spaces ( ~ ) LaTeX will not break
% a structure at a ~ so this keeps an author's name from being broken across
% two lines.
% use \thanks{} to gain access to the first footnote area
% a separate \thanks must be used for each paragraph as LaTeX2e's \thanks
% was not built to handle multiple paragraphs
%
\author{Ashiyana~Abdul Majeed,
        Mahmoud~Meribout,~\IEEEmembership{Senior Member,~IEEE,}
        and~Safa~Mohammed Sali % <-this % stops a space
\thanks{Ashiyana Abdul Majeed, Dr Mahmoud Meribout, and Safa Mohammed Sali are with the Department
of Computer and Information Engineering, Khalifa University, Abu Dhabi, UAE (email: 100059454@ku.ac.ae, mahmoud.meribout@ku.ac.ae, safa.msali@ku.ac.ae).}}
\maketitle

% As a general rule, do not put math, special symbols or citations
% in the abstract or keywords.
\begin{abstract}
In recent years, the development of specialized edge computing devices has significantly increased, driven by the growing demand for AI models. These devices, such as the NVIDIA Jetson series, must efficiently handle increased data processing and storage requirements. However, despite these advancements, there remains a lack of frameworks that automate the optimal execution of \ac{DNN}. Therefore, efforts have been made to create schedulers that can manage complex data processing needs while ensuring the efficient utilization of all available accelerators within these devices, including the CPU, GPU, \ac{DLA}, \ac{PVA}, and \ac{VIC}. Such schedulers would maximize the performance of edge computing systems, which is crucial in resource-constrained environments. This paper aims to comprehensively review the various \ac{DNN} schedulers implemented on NVIDIA Jetson devices. It examines their methodologies, performance, and effectiveness in addressing the demands of modern AI workloads. By analyzing these schedulers, this review highlights the current state of the research in the field. It identifies future research and development areas, further enhancing edge computing devices' capabilities.
\end{abstract}

% Note that keywords are not normally used for peerreview papers.

% For peer review papers, you can put extra information on the cover
% page as needed:
% \ifCLASSOPTIONpeerreview
% \begin{center} \bfseries EDICS Category: 3-BBND \end{center}
% \fi
%
% For peerreview papers, this IEEEtran command inserts a page break and
% creates the second title. It will be ignored for other modes.
\IEEEpeerreviewmaketitle

\vspace{-0.5em}
\section{Introduction}
% The very first letter is a 2 line initial drop letter followed
% by the rest of the first word in caps.
% 
% form to use if the first word consists of a single letter:
% \IEEEPARstart{A}{demo} file is ....
% 
% form to use if you need the single drop letter followed by
% normal text (unknown if ever used by the IEEE):
% \IEEEPARstart{A}{}demo file is ....
% 
% Some journals put the first two words in caps:
% \IEEEPARstart{T}{his demo} file is ....
% 
% Here we have the typical use of a "T" for an initial drop letter
% and "HIS" in caps to complete the first word.
\IEEEPARstart{A}{s} \ac{AI-SoC} become increasingly popular and widely adopted in various embedded devices, it is vital to allocate the tasks in such a way that ensures high performance while operating under low power, particularly in edge devices \cite{Kang-2020}. One such powerful edge device is the NVIDIA Jetson series, which contains hardware accelerators that are well-adapted for AI and graphics applications. They comprise several hardware engines dedicated to various parallel computation models and interfaced with high-speed and low-power \ac{SDRAM}. However, in most cases, these devices balance heavy workloads, including multiple \ac{DNN}s, that can impair their performance. Some works consider optimizing the \ac{DNN} execution in their GPUs, such as \cite{GPU-Support}, but often overlook other accelerators, leading to suboptimal performance. Proper task scheduling and partitioning into their hardware engines can significantly enhance efficiency, thereby using the full potential of edge computing. Building schedulers remains challenging due to the heterogeneous nature of their hardware architecture and the unpredictable memory contention between different hardware engines. The contention arises from using a shared bus that connects all hardware engines to the shared \ac{SDRAM}. Hence, extensive work is necessary to develop these devices' scheduling and hardware partitioning algorithms.   

With the rising demand for Jetson devices, it is of the utmost importance that a framework is developed to guarantee efficient and effective performance. For instance, the Jetson series has been used in autonomous driving. \cite{Pedestrian_Detection} evaluates the use of the Xavier and Orin device for pedestrian detection using an MM-Net model. Another example is a delivery robot that uses the Xavier device to navigate and reach the destination. Here, a modified \ac{SSD} is utilized for object detection \cite{Delivery-Robot}. Apart from robotics, the Jetson series has also been utilized for monitoring purposes, particularly in oceanography. Employing such devices allows researchers to automate the process safely and more economically, as opposed to traditional methods of physically measuring the data. It also gives them access to real-time data and insights \cite{nvidiaSaildroneCharts}. Moreover, it has also been used in industrial inspection to recognize defects \cite{Industrial_Inspection}. 

Implementing a scheduler would offer a more efficient, better-utilized system in each application. Ideally, the schedulers maximize the throughput while minimizing latency and power consumption. However, satisfying these goals simultaneously is not trivial, so most algorithms tend to optimize only a few of them, while keeping the remaining above a threshold value. Taking the case of the delivery robot \cite{Delivery-Robot}, implementing a \ac{HaX-CoNN} scheduler would enable multiple \ac{DNN}, each focused on different tasks to be concurrently executed. For instance, the \ac{SSD} was trained to identify people so another \ac{DNN} like ResNet can detect the other obstacles. Allocating specific roles to \ac{DNN} can improve the system's performance rather than using a single, more generalized model. Moreover, using \ac{HaX-CoNN} can improve latency by 20\% over GPU-only execution, thereby increasing the efficiency of the system \cite{HaX-CoNN}. An alternative scheduler that could be utilized is the \ac{CP-CNN} scheduler, which allows the concurrent processing of frames for a single \ac{DNN} model. \ac{CP-CNN} can enhance the speed and efficiency of the \ac{SSD} model by 75.6\% and 75.9\% respectively \cite{CP-CNN}. As AI models increase in complexity, it is vital to prioritize energy and latency in their performance. In fact, with the rise in the use of \ac{LLM}, there has been a renewed interest in the energy-efficient execution of these models on the edge \cite{LLM}. 

To date, no review paper has comprehensively compared the various schedulers for the recent NVIDIA Jetson edge accelerators. In \cite{Survey_Jetson_Old}, a survey on hardware and algorithmic optimizations for the old Jetson models was examined, namely the Jetson TK1, TX1, and TX2, where a \ac{DLA} is absent. These algorithmic optimizations were limited to modifying the characteristics of the \ac{DNN} used, including reducing the size of the \ac{DNN}, image resolution, number of frames, and depth-map levels. Additionally, the same survey explored works that did consider leveraging both the GPU and CPU present in the Jetson devices to improve the performance of the \ac{DNN} models, covering similar issues to those discussed in this paper, such as memory contention. Another review paper that dived into scheduling is \cite{Survey_GPU_Data_Centers}, which considers training and inference workload scheduling for GPU data centers. It is similar to edge computing in terms of accelerator heterogeneity, as it contains CPU and GPU resources. Furthermore, there is also a resemblance in the approaches used for deep learning model training, where linear programming and performance modeling are popular in scheduling training jobs. This paper seeks to systematically compare various schedulers and their methodologies targeting more recent NVIDIA edge devices while also identifying areas for potential development. 

% You must have at least 2 lines in the paragraph with the drop letter
% (should never be an issue)

\section{Background}
\subsection{Selection Criteria}
The papers were primarily conference and journal papers published in ACM and IEEE publications. Publications available on Google Scholar that contained the terms \enquote{NVIDIA Jetson AGX Orin} or \enquote{NVIDIA Jetson AGX Xavier}, \enquote{\ac{DNN}} or \enquote{CNN}, \enquote{\ac{DLA}} and \enquote{scheduling} were included in the review. Papers discussing hybrid edge-cloud systems were excluded from the review. While schedulers have been developed for other systems like the Google Coral devices \cite{RESPECT}, they are not truly heterogeneous as the GPU is used for graphical tasks, not for \ac{DNN} inference. In the case of the Libre devices \cite{Libre}, scheduling algorithms have not been developed as of now. Moreover, these devices target lightweight applications. Therefore, the Jetson devices were chosen as a representative hardware platform for this review, as they embody the defining characteristics of modern heterogeneous architectures, comprising of the CPU, GPU, and dedicated AI accelerators. Focusing on a single representative platform ensures a fair comparison of schedulers on a consistent hardware baseline. Additionally, the scheduling techniques discussed in this review can be extended to other comparable hardware platforms. Thus, the insights derived from these studies on the Jetson devices are indicative of broader scheduling behaviors in heterogeneous edge systems.

\begin{figure}[H]
    \centering
    \includegraphics[width=\linewidth]{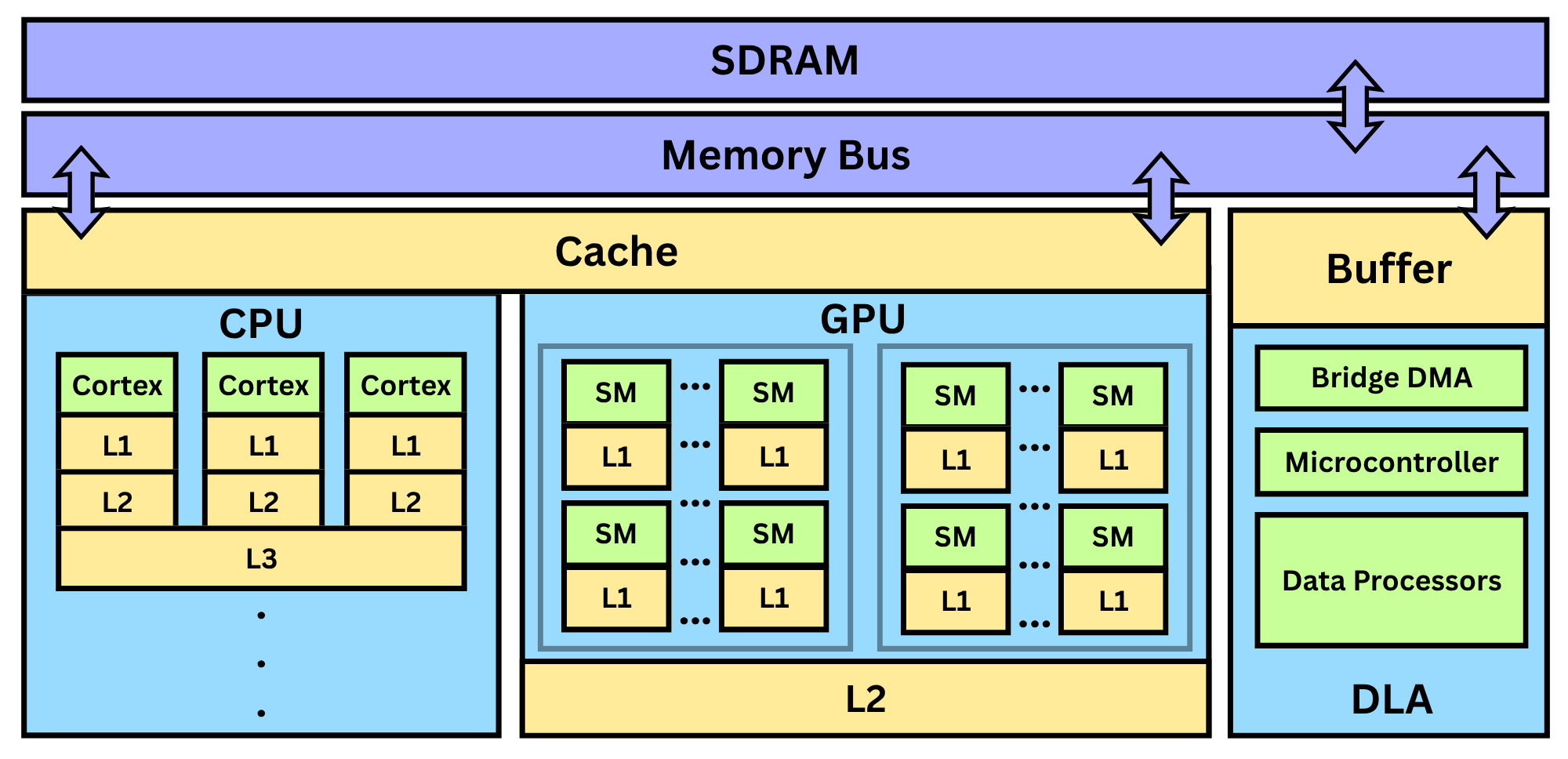}
    \caption{ Block diagram of the NVIDIA Jetson devices}
    \label{fig:Orin_Architecture}
\end{figure}

\subsection{Hardware Architecture of the Jetson AGX Devices}

The NVIDIA Jetson AGX devices (Fig. \ref{fig:Orin_Architecture}) follow the general architecture of heterogeneous edge GPUs , consisting of the CPU, GPU, \ac{DLA} (AI accelerator), and additional processors (like \ac{PVA} and \ac{VIC}). Compared to Xavier, the NVIDIA AGX Orin can provide almost 8 times the TOPS for AI and deep learning applications. For instance, using PeopleNet, the Orin device can produce 536 inferences per second, whereas Xavier can only output 196 inferences in the same period \cite{orin-techincal-brief}. 

\subsubsection{\textbf{CPU}}
The Orin device consists of 3 sets of 4 Cortex-A78 chips \cite{orin-techincal-brief} instead of a Carmel CPU present in Xavier \cite{xavier-orin-data}. Apart from the three levels of cache, the system also includes a 4 MB system cache. The maximum operating frequency is 2.2 GHz \cite{orin-techincal-brief}. It communicates with the memory using the system coherency fabric \cite{xavier-orin-data}. 

\subsubsection{\textbf{GPU}}
The Orin device uses an Ampere GPU \cite{orin-techincal-brief} (rather than Volta in Xavier \cite{xavier-orin-data}), operating at a maximum frequency of 1.3 GHz. There are 16 \ac{SM} and two levels of cache present in the device \cite{orin-techincal-brief}. It is connected to the memory using the memory subsystem interface. The GPU module alone can yield up to 170 Sparse TOPS to run AI models \cite{xavier-orin-data}. 

\subsubsection{\textbf{\ac{DLA}}}
The \ac{DLA} is essentially a fixed-function accelerator whose computational power is not comparable to GPU/CUDA but offers significantly higher energy efficiency. 2MB of \ac{SDRAM} is dedicated for use by the \ac{DLA} \cite{nvdla-doc}. The Orin device houses a more recent version of the \ac{DLA} than the Xavier device, increasing the performance by a factor of 9. This improvement is attributed to its architecture, where a local buffer is introduced, and \ac{SDRAM} bandwidth is reduced. The \ac{DLA} houses specialized units such as data processors, convolution cores, reshape engines, and additional processing elements \cite{orin-techincal-brief}. 

Apart from these hardware elements, the Jetson devices integrate two additional accelerators: the \ac{PVA} and \ac{VIC}. The \ac{PVA} consists of a Cortex-R5 chip, two \ac{DMA} engines, and two \ac{VLIW} \ac{VPU}. The Orin version includes 1MB of L2 memory cache \cite{orin-techincal-brief}. It can handle relatively light, post-processing tasks like feature detection, stereo disparity, feature tracking, and object tracking \cite{xavier-orin-data}. \ac{VIC} is responsible for pre-processing techniques like lens distortion correction, temporal noise reduction, and pixel processing \cite{xavier-orin-data}, operating up to 729.6 MHz on Orin and 1036.8 MHz on Xavier \cite{VIC_and_power}. To our knowledge, no paper has used the \ac{PVA} and \ac{VIC} modules when scheduling tasks. Only \cite{MEPHESTO} models the energy and memory contention between the CPU, GPU, and \ac{PVA}, and its approach could be extended to aid the scheduling algorithm in selecting the best execution pipeline. The \ac{PVA} could be used for convolution layers, and the \ac{VIC} could tackle pre-processing steps.

\vspace{-0.7em}
\subsection{Software Development Kits and Libraries}
NVIDIA provides resources allowing users to access the different modules in the Jetson devices and execute tasks on the desired module. CUDA \cite{CUDA} and cuBLAS \cite{cuBLAS} libraries are still necessary for more complex parallelism in the multi-core system. 
\subsubsection{\textbf{TensorRT}} It allows users to implement trained \ac{DNN} models on the NVIDIA Jetson devices by building an optimized inference engine. It enables users to alter techniques used in execution, such as mixed precision, and the option to choose the device for execution: GPU or \ac{DLA}. TensorRT also offers inter-layer and intra-layer optimization techniques to ensure the best engine performance \cite{tensorrt}. 
\subsubsection{\textbf{DeepStream}}  It is built on top of other libraries, like TensorRT, to assist in developing AI applications. It can construct a complete setup from input to output for use in computer vision applications \cite{deepstream}. 
\subsubsection{\textbf{\ac{VPI}}}  It is a library that enables the use of image and video processing algorithms, allowing access to \ac{PVA} and \ac{VIC}. Some examples include image conversion, filtering, lens distortion correction, and feature tracking. It is part of the DeepStream SDK \cite{VPI_documentation}. 

When executing tasks incompatible with the accelerator, the above resources redirect the task to be executed on the GPU. This can lead to interruptions if a different task already occupies the GPU. Furthermore, TensorRT does not allow for \ac{DNN} convolution layers to be assigned to the \ac{PVA}, limiting the use of the \ac{PVA} for filtering and other pre-processing tasks. Other libraries like the OpenVINO, Tensorflow, and ONNX may also be used to implement \ac{DNN} models on the NVIDIA Jetson edge devices. However, they come with a disadvantage as the DLA requires a TensorRT engine plan to utilize the device.
\vspace{-0.5em}
\subsection{Challenges of Building a Scheduler}
When building a scheduler, certain challenges need to be addressed, which are listed below: 
\subsubsection{\textbf{Memory contention}} 
Memory contention occurs not only in the \ac{SDRAM} and secondary memory but also within the cache and between the accelerators. As shown in Fig. \ref{fig:Orin_Architecture}, the GPU and specialized accelerators all share a common memory bus, the memory controller fabric, connected to an external LDDR5 \ac{SDRAM}. Such a system necessitates careful management to ensure no conflict arises from accelerators simultaneously accessing this memory \cite{HaX-CoNN, PCCS}. A similar problem of a single memory bus is also present in edge \ac{TPU}. The elements must wait if parameters are not cached, increasing memory traffic and power consumption. When executing AI models, almost 50.3\% of energy is spent on off-chip traffic \cite{Mensa}. Effective scheduling must account for these potential bottlenecks to optimize overall system performance.

\subsubsection{\textbf{Transition cost}} 
When allocating layers to the appropriate accelerators, it is also vital to consider the possible latency incurred if consecutive layers are assigned to different devices. Data from the previous accelerator's memory must be transferred to the shared memory for the next accelerator to access the data. Hence, a latency is introduced, impacting the overall performance \cite{HaX-CoNN}. As such, running the consecutive layer on the same device in certain stages of the \ac{DNN} may be more effective in avoiding the increasing latency. Thus, a trade-off needs to be made between the total transition time and the number of transitions that can occur to avoid compromising the performance of the \ac{DNN} model.

\subsubsection{\textbf{Accelerator properties}}
Each accelerator has specific limitations concerning the tasks or layers it can handle and the input/output formats it can accept. For example, depending on the function, the \ac{PVA} and \ac{VIC} are restricted to particular formats. The \ac{PVA} can implement convolution and filter operations on 8-bit and 16-bit unsigned integer formats. In contrast, the \ac{VIC} can implement image conversions and other manipulations on the same formats, usually on luma versions. The \ac{PVA}'s performance is close to or even faster than the CPU's for convolution. For instance, an unsigned integer 8-bit data of size 1920 by 1080 and a kernel of size three, the \ac{PVA} takes 0.27 ms, whereas the CPU takes 0.297 ms \cite{VPI_documentation}. The performance of \ac{VIC} is faster than the CPU for rescaling and some remapping tasks. In rescaling a 640 by 480 image to 1920 by 1080, the \ac{VIC} takes around 0.893 ms compared to the CPU, which takes about 6.75 ms \cite{VPI_documentation}. When looking at the \ac{DLA} in particular, similar constraints are present \cite{tensorrt}. The \ac{DLA} only supports FP16 and INT8 formats, with certain operations restricted to one of these formats. For example, the Equal operation is supported by INT8 alone. With \ac{DNN} layers, kernel sizes must range from 1 to 32, and the padding in deconvolution layers must be zero. In most cases, to ensure that the \ac{DLA} can execute the portion of the \ac{DNN} model, most schedulers \cite{AxoNN, CP-CNN, Jedi} will allocate the beginning of the \ac{DNN} model to the \ac{DLA} as these layers tend to be the most compatible rather than the end of the \ac{DNN} model. Hence, it is crucial to carefully consider layer specifications, input format, output format, and hardware limitations when assigning each layer to a device. This approach ensures compatibility and maximizes processing efficiency. 

\subsubsection{\textbf{Static versus Dynamic Schedulers}} 
Schedulers can be distinguished based on parallelization techniques as well as scheduling methods. Static schedulers follow a predetermined plan, whereas dynamic schedulers actively adjust the execution strategy at run-time based on the performance metrics or the specific objectives defined by the algorithm. The aim could be minimizing latency, maximizing throughput, or reducing energy consumption, depending on the scheduler. While static schedulers are more straightforward to implement, they are inflexible when sudden changes occur during the execution pipeline. In such cases, dynamic schedulers are preferred. Almost all the schedulers referenced in this paper are static, except \ac{D-HaX-CoNN} \cite{HaX-CoNN} and \ac{H2H} \cite{H2H}. In the case of edge AI applications, most have a deterministic pipeline that will continue execution. Therefore, a static scheduler is sufficient in most applications. Dynamic schedulers can be implemented in the few cases where the pipeline is unspecified, particularly when different models are required for various phases, like \cite{Zhuyi}.

\subsubsection{\textbf{Horizontal versus Vertical Schedulers}} 
In most cases, schedulers divide the layers in the \ac{DNN} model into sub-groups to be executed on the designated accelerator, called vertical pipelining. This can be ideal for the parallel processing of video frames and for multi-\ac{DNN} execution. On the other hand, few schedulers consider dividing the \ac{DNN} based on the width dimension so that each accelerator can independently execute a subset of the \ac{DNN} model. This type of execution could be preferable for cases with a single \ac{DNN} model. 

\subsection{Metrics}
The performance of schedulers is assessed using three key metrics: throughput (\ensuremath{T}), latency (\ensuremath{L}), and energy (\ensuremath{E}). The throughput refers to the amount of data a \ac{DNN} model can process within a specific time frame, while latency refers to the time required to process a single input. To aid in identifying the ideal schedule, utilization may be used as it evaluates the extent to which the processing power of the available accelerators is being employed. Various other values have been used in scheduling problem formulation, as defined below:
\begin{itemize}
    \item \( a_i \) represents a single task from a set \( A \) of all three \ac{DNN} tasks and three \ac{RNN} tasks present in the Apollo autonomous driving software \cite{Apollo} (from \cite{LP}).
    \item \( ce_j \) represents a processing element from a set \( CE \) of all possible processing elements, which include 2, 4, or 6 CPU cores, 4 or 8 GPU cores, and finally 1 or 2 \ac{DLA} units (from \cite{LP}). 
    \item \( L(a_i, ce_j) \) represents the latency of a task \( a_i \) executed on the element \( ce_j \) (from \cite{LP}).
    \item \( B(a_i, ce_j) \) represents the binary decision as to whether or not the task \( a_i \) is executed on the element \( ce_j \) (from \cite{LP}).
    \item \( N \) represents a \ac{DNN} model and \( N_n \) represents a layer \( n \) present in a \ac{DNN} \( N \) (from \cite{AxoNN,HaX-CoNN}). 
    \item \( s(N_n) \) represents the hardware element on which the layer \( N_n \) is executed (from \cite{AxoNN,HaX-CoNN}).
    \item \( L(N_n, s(N_n)) \) represents the latency of a layer \( N_n \) executed on the hardware element \( s(N_n) \) (from \cite{AxoNN,HaX-CoNN}).
    \item \( TR_n \) represents the binary decision on whether a transition occurs at this layer \( N_n \) (from \cite{AxoNN,HaX-CoNN}). 
    \item \( \tau(N_n, s(N_n), \text{OUT}) \) and \( \tau(N_{n+1}, s(N_{n+1}), \text{IN}) \) represents the transition cost (time) for the layer \( N_n \) to exit the device \( s(N_n) \) and for the layer \( N_{n+1} \) to enter the device \( s(N_{n+1}) \) (from \cite{AxoNN,HaX-CoNN}). 
    \item \( pipeline(N_n, s(N_n)) \)  represents the cost involved when the sub-unit executing the layer \( N_n \) is not the same as the previous layer \( N_{n+1} \) (from \cite{AxoNN}). 
    \item \( e(N_n, s(N_n)) \) represents the energy consumed when executing layer \( N_n \) on the device \( s(N_n) \). Similarly, the \( e(N_n, s(N_n), \text{OUT}) \) and \( e(N_{n+1}, s(N_{n+1}), \text{IN}) \) represents energy consumed for the layer \( N_n \) to exit the device \( s(N_n) \) and for the layer \( N_{n+1} \) to enter the device \( s(N_{n+1}) \) (from \cite{AxoNN}).
    \item \( C_{N_n, s(N_n)} \) represents the slowdown on layer \( N_n \) resulting from contention during the concurrent execution of layers (from \cite{HaX-CoNN}).
    \item \( ECT \) represents user's energy consumption target (from \cite{AxoNN,Map_&_Conquer}).
    \item \( LT \) represents user's latency target (from \cite{Map_&_Conquer}).    
    \item \( P \) represents user's performance objective (from \cite{Map_&_Conquer}).
    \item \( size(F, I) \) represents memory required for storing the immediate features from each inference stage, \( F \), and the indicator matrix, \( I \), which states whether these features are required in the following stage (from \cite{Map_&_Conquer}). 
    \item \( M \) represents the maximum shared memory available to the processing elements (from \cite{Map_&_Conquer}). 
\end{itemize}

These variables used in problem formulation have been curated from schedulers discussed in the upcoming sections. These were directly adopted from corresponding references.

\section{Schedulers}
Schedulers developed for the NVIDIA edge devices can be categorized based on the methodology they use to find the optimal pipeline. They either use an optimization technique or a heuristic method to arrive at the solution. Heuristic schedulers reach the near-optimal configuration through an algorithm to find the best partitioning solution using a set of metrics that could involve latency, throughput, and power consumption. They depend on intuitive or experience-driven strategies to model the problem. As such, the optimal solution may only approximate the final schedule. On the other hand, optimization methods rely on formal techniques, like linear programming and satisfiability problem solvers, to determine the optimal schedule that fulfills the objective function. The constraints emulate the limitations of the device and the application. Most of the schedulers have been developed for the Xavier device. Since both the Xavier and Orin devices are comparable architecturally, the schedulers are expected to function similarly. 
\vspace{-0.5em}
\subsection{Heuristic-Based Scheduling Techniques}
\subsubsection{\textbf{Jedi}}
In \ac{Jedi} \cite{Jedi}, the solution is finalized when the algorithm reaches the highest throughput while maintaining a satisfactory utilization. This is done through a heuristic global search first, where a sample cut-point set allows the algorithm to locate an estimated cut-point, followed by a heuristic local search to determine the exact location of the cut-point. Additionally, experiments were conducted to attain the optimal parameter combination for increasing throughput. The first parameter is the number of threads for the operation of pre-processing and post-processing tasks, where they are allowed to run concurrently to prevent bottlenecks from occurring. The number of threads ranges from 1 to 6. The second parameter is the number of streams of the assigned kernel that can be run concurrently without performance saturation occurring. Using the TensorRT framework, the number of streams ranges from 1 to 6. The third parameter is the number of stages the pipeline can be divided into for running on different devices. This can be GPU-only, \ac{DLA}-GPU, or GPU-\ac{DLA}-GPU. The fourth and final parameter is the application of \ac{PND}, where the network portion is duplicated on the two \ac{DLA} units and runs concurrently in the same stage, increasing the parallelism and utilization of the accelerators. 

These parameters are implemented in the overall pipeline to produce an execution timeline shown by Fig. \ref{fig:Timing_Diagram}. In most cases, the number of streams for pre-processing varies from 1 to 6, and for post-processing, it ranges from 1 to 3. The pipeline used the \ac{DLA} with GPU in most cases, along with \ac{PND} to use both the \ac{DLA} units. A GPU alone could infer smaller models, like YOLOv3 tiny in integer 8-bit format. Within the accelerator, the \ac{DLA} implemented four streams while the GPU implemented two.

\subsubsection{\textbf{CP-CNN}}

\ac{CP-CNN} \cite{CP-CNN} uses a different approach where the latency and computing power are the deciding factors in finding the partitioning point. The objective of \ac{CP-CNN} is to ensure that the execution time in the GPU is equivalent to the execution time in the \ac{DLA}, as shown in Fig. \ref{fig:Timing_Diagram}. The latency of the set of frames is reduced under \ac{CP-CNN} with the implementation of the \ac{DLA} and the concurrent processing. Since the \ac{DLA} latency and GPU latency are equal, at no point in time are the \ac{DLA} and GPU left idle.

The \ac{CP-CNN} algorithm iteratively compares the ratio of operations from the first layer to the current layer to the total operations in all layers (operation ratio) with the ratio of the computing power of the \ac{DLA} to the total computing power (computing power ratio). The partitioning point is set as the current layer if the operation ratio is less than or equal to the computing power ratio. The algorithm also simultaneously compares the total time (operation time on \ac{DLA} in addition to the data transfer time) required to execute these layers. If the latency to run till the partitioning point is more than the latency to run till the partitioning point with one additional layer, the partitioning point is incremented by one \cite{CP-CNN}. 

\subsubsection{\textbf{Herald and H2H}}
\begin{figure} [H]
    \centering
    \includegraphics[width=\linewidth]{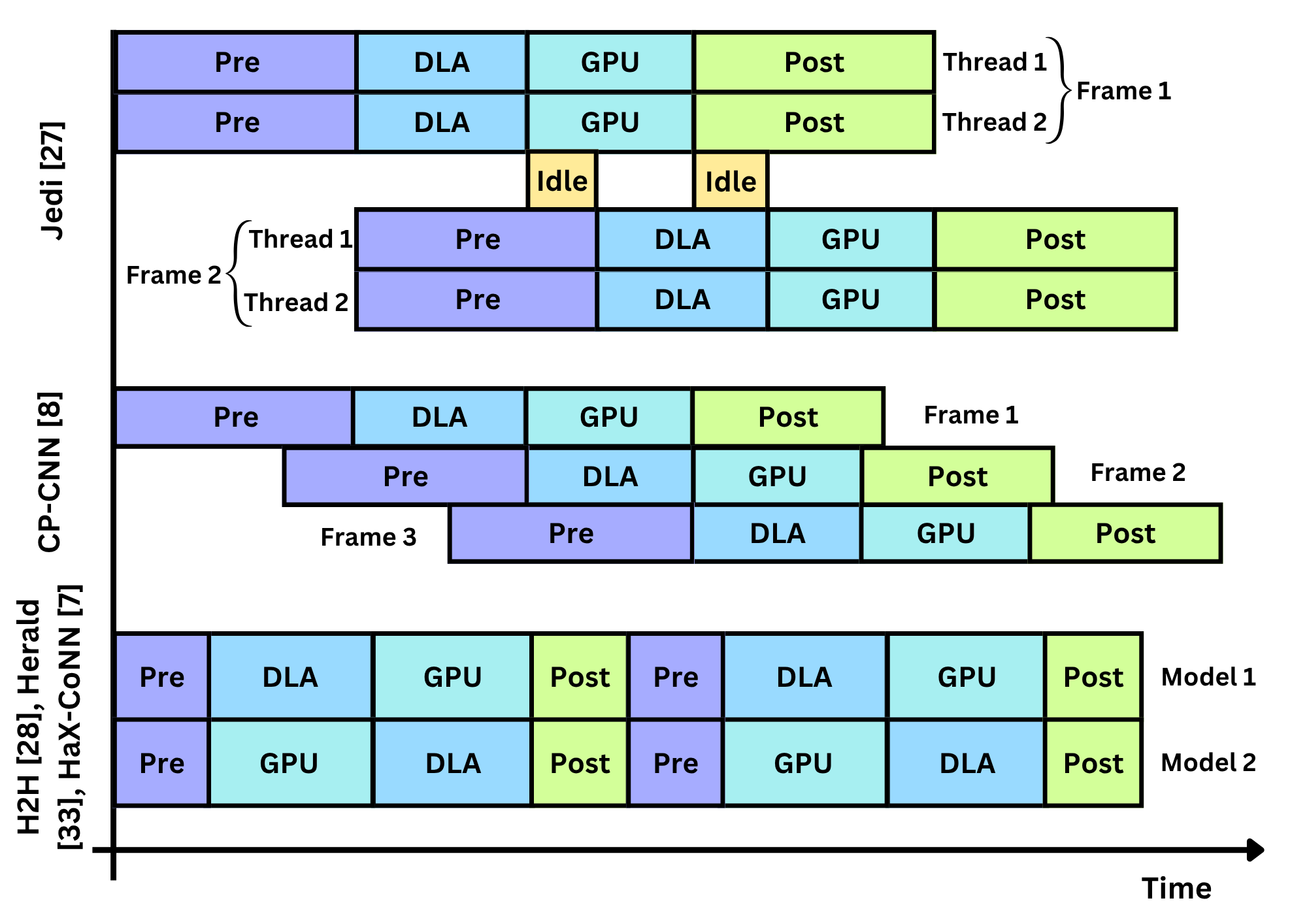}
    \caption{ Timing diagram of \ac{Jedi}, \ac{CP-CNN}, and multiple model schedulers (includes \ac{H2H}, Herald, and \ac{HaX-CoNN})}
    \label{fig:Timing_Diagram}
\end{figure}

Much like \ac{CP-CNN} and \ac{Jedi} algorithms, Herald \cite{Herald} divides the layers and assigns them to the processing elements, but also considers the concurrent execution of \ac{DNN}s. It implements a switching technique between the \ac{DLA} and GPU execution, as depicted in Fig. \ref{fig:Timing_Diagram}, which results in a swifter pipeline for the concurrent execution of \ac{DNN}s.

A design-space exploration algorithm is implemented to assign layers based on data flow, following which post-processing is done to eliminate idle time. The \ac{H2H} \cite{H2H} builds on Herald while including transition costs in the layer assignment process. After performance modeling, the mapping algorithm follows four steps to complete the scheduling task. First, an initial mapping is generated based on assessing the layer-wise computation, while ignoring all other factors. This mapping buffers weights into the device's local \ac{SDRAM} using a Knapsack algorithm. Adjacent layers are merged to minimize further the latency incurred from the data transfer of input/output feature maps. Finally, the scheduler tries to compromise between transition and computation costs, resulting in a more transition-cost aware scheme. The dynamic version of the \ac{H2H} scheduler uses a modified Knapsack algorithm \cite{H2H}.

\subsection{Optimization-Based Scheduling Techniques}
\subsubsection{\textbf{LP}}
In contrast to most methods listed \cite{AxoNN, CP-CNN, HaX-CoNN, Jedi, Kang-2020}, \ac{LP} methodology \cite{LP} maps the entire \ac{DNN} instance onto a processing element, which can either refer to the sub-units or the whole accelerator (CPU, GPU, or \ac{DLA}). Since the \ac{DNN} as a whole is executed on the processing element, transition points between layers are not considered. Within the multi-core CPU module, 2, 4, or 6 cores are considered using the OpenMP library \cite{OpenMP}. For the GPU module, 4 or 8 \ac{SM} are considered using the CUDA and cuBLAS library \cite{cuBLAS, CUDA}. Using TensorRT \cite{tensorrt}, 1 or 2 units of the \ac{DLA} are considered. The \ac{LP} scheduler is built for use with the Apollo Autonomous Driving software \cite{Apollo}, consisting of perception, speech recognition, trajectory planning, localization, mapping, prediction, and control modules. The perception module has four \ac{DNN}s for different object detection and tracking purposes. One of them is excluded from scheduling as it requires more resources than what is available from the Xavier device. The speech recognition module also uses a \ac{DNN}. The prediction module has three \ac{RNN} models for lane sequence computation. These \ac{DNN}s and \ac{RNN}s are considered for scheduling and were evaluated to check for compatibility with the accelerators, which clarifies that \ac{RNN}s cannot be executed on \ac{DLA}, and some \ac{DNN}s cannot be executed on the CPU. The \ac{DNN}s and \ac{RNN}s are executed simultaneously in the Xavier device, where each neural network instance is executed recurrently based on a defined frame rate, making the scheduler static. The objective function of the \ac{LP} scheduler focuses on minimizing the cumulative latency of the system (Equation \ref{eq:LP_equation}).

{
\setlength{\abovedisplayskip}{5pt}%
\setlength{\belowdisplayskip}{5pt}%
\begingroup
\footnotesize \begin{equation} \label{eq:LP_equation}
\min \sum_{a_i \in A, \, ce_j \in CE} L(a_i, ce_j) \cdot B(a_i, ce_j)
\end{equation}
\endgroup
}

The latency of each task was found through profiling on each device \cite{LP}. Memory contention was considered an assumption with a time factor added to the profiled timings. It does not consider the system's energy consumption. The constraints include ensuring that a task \( a_i \) can only be mapped to a single processing element \( ce_j \) and that when a \( ce_j \) is chosen, some of the other values in the set \( CE \) become obsolete. For example, if both the \ac{DLA} units are chosen for a single \ac{DNN} instance, the individual \ac{DLA} options are no longer considered for the remaining \( a_i \) present in  \( A \). Similarly, in the case of GPU, if all 8 \ac{SM} are considered for a particular model, the 4 \ac{SM} option is no longer present in the set \( CE \). Rather than presenting the throughput of the system during the concurrent processing, the scheduler is assessed on its success rate (feasibility ratio) of executing the \ac{DNN}s and \ac{RNN}s on the set of \( CE \). The results demonstrated that adding more devices to the set \( CE \) increases the feasibility ratio of scheduling increased workloads \cite{LP}.

\subsubsection{\textbf{AxoNN}}
The \ac{AxoNN} scheduler \cite{AxoNN} follows a similar methodology to \ac{CP-CNN}, dividing layers into \ac{DLA} then GPU. \ac{AxoNN} uses a constraint-based optimization problem focused on satisfiability along with empirical modeling. The energy, latency, pipeline, and transition times are found through performance modeling. A Z3 \ac{SMT} solver is utilized to identify the ideal pipeline. The same solver has also been used in the \ac{HaX-CoNN} scheduler for a \ac{SAT} problem \cite{HaX-CoNN}. The objective function (Equation \ref{eq:AxoNN_objective}) aims to minimize the latency while ensuring that the energy consumed is below the threshold value set by the user. Equations \ref{eq:AxoNN_equation_time} and \ref{eq:AxoNN_equation_energy} show how these values are calculated \cite{AxoNN}.%

{\setlength{\abovedisplayskip}{1pt}
\setlength{\belowdisplayskip}{1pt}
\begingroup
\footnotesize \begin{align} \label{eq:AxoNN_objective}
    \min \quad latency(N, CE, S(N \rightarrow CE)) \notag \\
    \mathrm{s.t.} \quad  energy(N, CE, S(N \rightarrow CE)) < ECT  
\end{align}
\endgroup}
\setlength{\abovedisplayskip}{1pt}
\setlength{\belowdisplayskip}{1pt}
\begingroup
\footnotesize \begin{align} \label{eq:AxoNN_equation_time}
    latency = \sum_{n=0}^{len(N)} 
    \Big( L(N_n, s(N_n)) + TR_n \cdot \tau(N_n, s(N_n), OUT)  \notag \\
    + TR_n \cdot \tau(N_{n+1}, s(N_{n+1}), IN) + TR_n \cdot pipeline(N_n, S(N_n)) \Big)
\end{align}
\endgroup% 
\setlength{\abovedisplayskip}{1pt}
\setlength{\belowdisplayskip}{5pt}
\begingroup
\footnotesize \begin{align} \label{eq:AxoNN_equation_energy}
    energy = \sum_{n=0}^{len(N)} 
    \Big( e(N_n, s(N_n)) + TR_n \cdot e(N_n, s(N_n), OUT)  \notag \\
    + TR_n \cdot e(N_{n+1}, s(N_{n+1}), IN) \Big)
\end{align}
\endgroup%

\ac{AxoNN} achieves an accuracy of 97.1\% on execution time prediction and 98.2\% on energy consumption prediction. In the worst-case scenario, the accuracy reduces to 78.1\% and 71.9\%, respectively. As mentioned previously, increasing the number of transitions not only involves an extra overhead of inter-layer performance but also requires more time for the solver to suggest the optimal scheduler, going from five seconds for a single transition to a minute for three transitions on the Jetson AGX Xavier \cite{AxoNN}. A comparison between the optimal schedule for a model and the GPU-only execution was absent from the paper, as this would vary depending on the \( ECT \) set by the user. 

\subsubsection{\textbf{GA}}
Another approach explored in \cite{Kang-2020} involved using a \ac{GA} to schedule tasks based on profiling data from the different accelerators. Similar to \ac{AxoNN} and \ac{HaX-CoNN}, the transition and execution time are profiled before it passes to the \ac{GA}. The \ac{GA} then generates the ideal partitioning point for the \ac{DNN}. The mechanism of the algorithm is similar to evolutionary processes in nature, where only the best-fit chromosome (an array of layers where the index represents the processing element) survives. The approach was initially tested on smartphones like the Galaxy S9, then later tested on the Xavier device in \cite{Jedi}. The objective of the \ac{GA} scheduler when using a single \ac{DNN} is to maximize the throughput and minimize the energy consumption. In the case of multiple \ac{DNN}s, the objective is to reduce the latency and energy consumption. The \ac{GA} method assigns layers of the \ac{DNN}s to the processing elements within the CPU, taking advantage of parallelism within and between the processing units.

\subsubsection{\textbf{HaX-CoNN}}
\ac{HaX-CoNN} \cite{HaX-CoNN} is built on Herald and \ac{H2H}. The difference is that it uses a constraint-based satisfiability problem to minimize the latency or maximize the throughput (the objective changes depending on the application), as shown by Equation \ref{eq:HaX-CoNN_objective}. Additionally, it considers memory contention caused by the concurrent execution of layers with the use of \ac{PCCS} \cite{PCCS}. This is evident from Equation \ref{eq:HaX-CoNN_equation_time}, where the value of \( C_{N_n, s(N_n)} \) is found using \ac{PCCS}. % 

{\setlength{\abovedisplayskip}{5pt}
\setlength{\belowdisplayskip}{1pt}
\begingroup
\footnotesize \begin{align} \label{eq:HaX-CoNN_objective}
    \max \sum_{i=0}^{len(DNN)} \frac{1}{latency(N, CE, S(N \rightarrow CE))_i} \notag \\
    \min \quad \max latency(N, CE, S(N \rightarrow CE))_i 
\end{align}
\endgroup}
{\setlength{\abovedisplayskip}{1pt}
\setlength{\belowdisplayskip}{5pt}
\begingroup
\footnotesize \begin{align} \label{eq:HaX-CoNN_equation_time}
    latency = \sum_{n=0}^{len(N)} 
    \Big( L(N_n, s(N_n)) \times C_{N_n, s(N_n)} \notag \\
    + TR_n \cdot \tau(N_{n+1}, s(N_{n+1}), IN) + TR_n \cdot \tau(N_n, s(N_n), OUT) \Big)
\end{align}
\endgroup}

\begin{figure}[H]
    \centering
    \includegraphics[width=\linewidth]{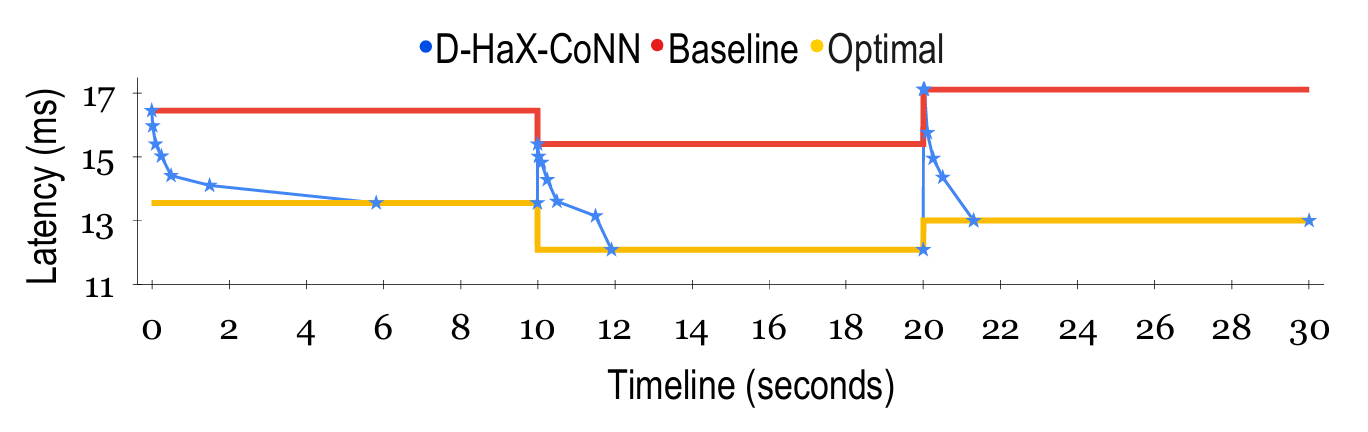}
    \caption{ Dynamic execution where the system changes every 10 seconds \cite{HaX-CoNN}}
    \label{fig:D-HaX-CoNN_result}
\end{figure}

Considering the application, \ac{HaX-CoNN} can suggest optimal schedules for various types of \ac{DNN} execution. For two instances of the same \ac{DNN} running concurrently, it improved throughput up to 29\% compared to GPU-only and GPU-\ac{DLA} execution. For two \ac{DNN}s running simultaneously on either the same data or streaming data, latency and throughput improved by 23\%. The CPU is not considered in executing tasks but rather to carry out the scheduling in the case of the dynamic version (\ac{D-HaX-CoNN}). \ac{D-HaX-CoNN} starts with an initial naïve schedule. With this, the concurrent \ac{DNN} loop begins execution, and its schedule is updated with a more optimized version by the Z3 solver. The solver runs until the schedule can no longer be optimized. Fig. \ref{fig:D-HaX-CoNN_result} shows the effectiveness of employing the \ac{D-HaX-CoNN} scheduler, where it can come to the ideal solution in less than two seconds for the two systems with a pair of \ac{DNN}s. There are a total of 3 changes in the graph. Part 1 is where GoogleNet and ResNet152 are run serially, while ResNet18 runs in parallel. Part 2 is where Inception and ResNet152 are run concurrently. Part 3 replaces Inception in the previous case with VGG19. In the case of the first part, the system has three \ac{DNN}s, and therefore more layer groups and possibilities to consider. Moreover, the overhead of running the Z3 solver on the CPU, alongside the concurrent \ac{DNN} execution in \ac{DLA} and GPU, is less than 2\% \cite{HaX-CoNN}. 

\subsubsection{\textbf{Map-and-Conquer}}
Map-and-Conquer \cite{Map_&_Conquer} follows a unique approach compared to all the schedulers mentioned above; firstly, it implements parallelism along the horizontal to divide the model into several independent inference stages that can run simultaneously, as opposed to the vertical pipelining (layer-wise). \cite{Hadidi-2020} argues that the division of layers across accelerators is not the most efficient way to utilize them. Operations, like convolution and multi-head self-attention, can be parallelized so that multiple operations of the same layer are concurrently carried out on different devices to maximize efficiency. Secondly, it is one of the first works to consider transformer models apart from \ac{DNN} models, making it more versatile in terms of model architecture. Similar to \ac{GA}, it uses an evolutionary algorithm to identify the ideal schedule, based on profiling data from TensorRT, and a predictor, XGBoost \cite{XGBoost}, to provide layer estimates of latency and energy on each of the devices to the algorithm. The predictor is required as it accounts for the computation and communication costs, similar to the \ac{PCCS} model in \ac{HaX-CoNN}. For VGG-19, compared to GPU-only execution, which consumes 630 mJ, Map-and-Conquer can provide a scheme with 136 mJ consumption (4.62x reduction) while keeping the latency relatively same (25 ms). The objective function of Map-and-Conquer is given by Equation \ref{eq:Map-and-Conquer_objective}. It allows the user to set an energy consumption threshold (similar to \ac{AxoNN}), latency threshold, and the performance objective, thus allowing for greater flexibility in the pipeline creation \cite{Map_&_Conquer}.

{\setlength{\abovedisplayskip}{1pt}
\begingroup
\footnotesize \begin{align} \label{eq:Map-and-Conquer_objective}
    \min \text{ } P \text{ } s.t. \text{ } latency < LT, energy < ECT, size(F, I) < M 
\end{align}
\endgroup
}

\subsubsection{\textbf{MaGNAS}}
The \ac{MaGNAS} is another scheduler that uses an evolutionary algorithm, but developed for vision \ac{GNN}. Compared to GPU-only execution, \ac{MaGNAS} is able to reduce latency by a factor of 1.57 and increase efficiency by a factor of 3.38 on the Xavier device. It utilizes a two-step evolutionary algorithm to optimize the architecture of the \ac{GNN} and perform the layer mapping based on the performance modeling. During architecture optimization, the different generations of architecture are evaluated on a fitness function that considers accuracy, latency, and energy. After ranking them, some are eliminated, and the remaining undergo mutation and crossover to get a new generation. The process continues until the search budget is reached. The exact process is carried out to identify the ideal hardware partitioning. Similar to Map-and-Conquer, it allows the user to set an energy consumption and latency threshold \cite{MaGNAS}. It does not consider the memory contention as part of the fitness function. 
\vspace{-0.5em}
\section{Comparative Analysis}%
\subsection{Idle Time}
\ac{CP-CNN} seeks to reduce the idle time between the \ac{DLA} and GPU, unlike \ac{Jedi}, leading to faster inferences and more energy savings in the long run for streaming data. Fig. \ref{fig:Timing_Diagram} illustrate this difference as \ac{Jedi} has idle instances in which either the \ac{DLA} or the GPU is unused. Herald and \ac{H2H} approach the scheduling of concurrent DNNs similarly by seeking to reduce the idle time when splitting the models between the accelerators. This results in a faster inference compared to independent model execution on the accelerators, as seen in \ac{LP}. Regardless, they are ineffective due to the absence of an optimization framework and omission of memory contention, which is why HaX-CoNN is able to outperform them. The difference in approaches impacts the partitioning point of the model. For instance, in \ac{Jedi}, YOLOv3 is split into \ac{DLA} and GPU at the \nth{57} layer, whereas in \ac{CP-CNN}, the split occurs at the \nth{16} layer.
\begin{figure*}[!t]
    \centering
    \includegraphics[width=\linewidth]{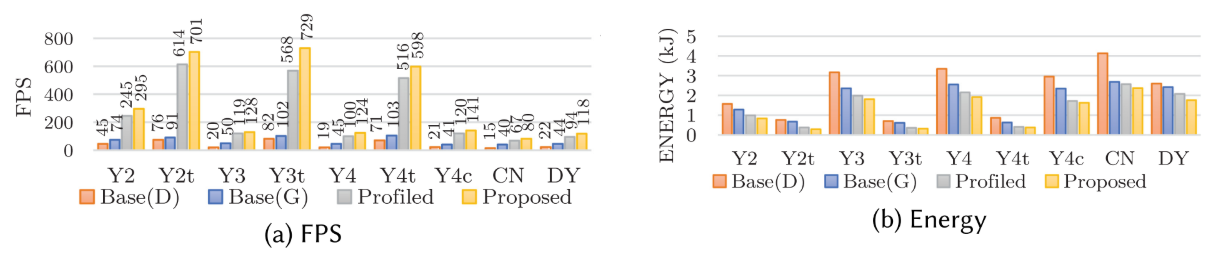}
    \caption{ Throughput and energy consumption of different \ac{DNN} models when executed on Jetson AGX Xavier using the schedulers. Y refers to the YOLO model, CN refers to CSPNet, and DY refers to DenseNet with YOLO. The performance of \ac{Jedi} (proposed), is compared with profiled (\ac{GA}) \cite{Kang-2020}, \ac{DLA} only (base(D)) and GPU only execution (base(G)). \cite{Jedi}}
    \label{fig:Jedi_graph}
\end{figure*}
\vspace{-0.7em}
\subsection{Intra-Accelerator Parallelism}
Among all the schedulers, \ac{Jedi} maximizes intra-parallelization to the highest degree using multiple threads, streams, and \ac{PND}. However, finding the ideal combination is not as apparent since it is determined through trial and error. Extensive profiling data must be collected before the final schedule can be decided, as it is not just the layers but also the pipeline, streams, and threads that vary. Only the cut-point is obtained through the algorithm. The experiments also prove that the nature of the model impacts these parameters, so each model requires customized pipelines for efficient execution. Nevertheless, it is worth exploring as utilizing the parallelism within and between the accelerators improves the throughput and reduces power consumption compared to \ac{GA}, as evidenced by Fig. \ref{fig:Jedi_graph}. Other schedulers that implement concurrent \ac{DNN}s, like Herald, \ac{H2H}, and \ac{HaX-CoNN}, could be modified to take advantage of the parallelism within the GPU and \ac{DLA} to improve the inference efficiency further. Rather than letting the remaining \ac{DLA} and \ac{SM}s in the GPU remain idle, adding them to the pipeline would be more advantageous.%
\vspace{-0.7em}
\subsection{Energy Consumption}
Map-and-Conquer is the ideal choice for energy-sensitive applications as it optimizes the execution of a single model using horizontal scheduling. It is also adaptable in handling multiple model architectures, provided that the layers are \ac{DLA} compatible. AxoNN may be used as a workaround for more complex models where \ac{DLA} compatibility may be an issue. However, in severe cases, the incompatibility may need to be addressed at the development stage using modifications. Although \ac{AxoNN} and \ac{MaGNAS} also allow users to set an energy threshold, it does not leverage parallelism as one model infers a single frame at a time. Moreover, \ac{MaGNAS} specifically caters to \ac{GNN} models. In the remaining schedulers, reduced energy consumption is either an indirect consequence or one of the objective functions of the scheduler. This approach is ill-suited for applications with strict power constraints. Case in point, even though \ac{CP-CNN} considers the computing power in its algorithm, it uses the power ratio, not the power value itself, to determine the cut-point. It does not ensure that the optimal pipeline consumes the least energy.%
\vspace{-0.7em}
\subsection{Transitions and Transition Costs}
Generally, most schedulers employ a single data transfer from the \ac{DLA} to the GPU, as frequent switches between these devices introduce inefficiencies. This has proven to be one of the most efficient methods over constant switching between \ac{DLA} and GPU. In rare instances, using more than one transition is preferred, like CSPNet in \ac{Jedi}'s case. Even \ac{AxoNN} and \ac{CP-CNN} explore the possibility of implementing more than a single transition. However, \ac{AxoNN} opts for a single transition as bottlenecks in the limited bandwidth cause an increased transition time. In \ac{CP-CNN}, multiple transitions are considered for devices containing more than one GPU. Map-and-Conquer and \ac{LP} are the only schedulers that do not consider layer partitioning. Map-and-Conquer uses horizontal scheduling so the partitioning is done along the model's width, resulting in subsets that can be independently executed. However, this only allows a single model or frame to be executed. On the other hand, \ac{LP} executes the entire model on the device, only accounting for the utilization and latency. Over time, such a system can incur latency, potentially violating the application's requirements, as it might require the output in a specific time frame. It is worth noting that transition costs were considered in schedulers developed after 2021, including \ac{H2H}, \ac{AxoNN}, \ac{CP-CNN}, and \ac{HaX-CoNN}.%
\vspace{-0.8em}
\subsection{Memory Contention}
The only scheduler that accounts for the memory is the \ac{HaX-CoNN} scheduler. Adding this parameter to the algorithm improves the system's efficiency and latency prediction. Memory contention is an issue even with a single model due to the concurrent execution of the GPU and \ac{DLA} with multiple frames. Some schedulers, like \ac{AxoNN} and \ac{MaGNAS}, may not require memory contention as there is no concurrency. Only transition costs need to be accounted for. So far, \ac{PCCS} is the only technique to model contention, so further studies need to be carried out.%
\vspace{-0.8em}
\subsection{\ac{DLA} Use and Layer Compatibility}
Most schedulers treat two \ac{DLA}s as a single unit, reducing inter-device transitions and optimizing speed. However, in the case of the \ac{Jedi} scheduler, a \ac{PND} was introduced, allowing two instances of the same network to run concurrently. The typical order of device operation places the \ac{DLA} first, followed by the GPU, due to the nature of the task. The \ac{DLA} is well-suited for early-stage tasks like feature extraction, while the GPU better handles later layers for specialized tasks like object detection or image recognition. The issue of layer compatibility of the \ac{DLA} is reflected in the schedulers as well, where \ac{Jedi} and \ac{CP-CNN} can avoid this issue by implementing the transition point in the early part of YOLO models, which are otherwise not fully compatible with \ac{DLA}. In the case of \ac{LP}, \ac{AxoNN}, and \ac{HaX-CoNN}, the profiling stage ensures that the incompatible layers are permanently assigned to the GPU. Since the model's subset is allocated to the DLA for Map-and-Conquer, incompatible layers may disrupt the pipeline and fallback to the GPU. As DNN models become increasingly complex, the layers present at the beginning of the model may not always be DLA compatible. This may include padding in the deconvolution layer, which is present in many CNN-based \ac{GAN} models. Therefore, designing a system that can tackle such issues is beneficial by substituting the layers with similar operations before deployment. Current frameworks, like DeepStream and TensorRT, act defensively as they do not handle these issues and instead offload the CNN models as is to the GPU, resulting in pipeline disruptions.%
\vspace{-0.8em}
\subsection{Scheduler Methodology and Latency}
Schedulers often use a looped structure to organize layer execution according to specific methodologies, with variations in the parameters used for decision-making. The \ac{Jedi} method is distinguished by its multi-policy approach, while other schedulers assess each layer individually or as a group to seek transition points. In contrast, the \ac{LP} method addresses the task as a whole, leveraging device heterogeneity by allocating different \ac{DNN}s to each device. 
Most papers do not state the time required to identify the ideal schedule. This data point may not be as significant with static schedulers since the schedule is generated before the execution cycle. However, this is necessary for dynamic scheduling as the overhead may negatively impact the performance of the models running in parallel. For \ac{HaX-CoNN}, this is clearly mentioned (Fig. \ref{fig:D-HaX-CoNN_result}). The overhead of the scheduler is about 2\% when running on the CPU core. These data points are absent for dynamic \ac{H2H}.%
\vspace{-0.8em}
\subsection{Performance Analysis}
Most schedulers were assessed on DNN models like YOLOv3, VGG-19, ResNet, and DenseNet. Such models continue to be used in industrial applications, where edge performance is critical due to limited storage and computational resources. For instance, ResNet models have been used to classify and analyze soil for agricultural purposes \cite{agriculture}. YOLOv3 models have been used in Internet of Things applications, including monitoring and smart camera use \cite{YOLOv3_case_study}.
Currently, the schedulers are evaluated on different models, and no standard model has been tested across all of them. Additional experimentation is required to enable consistent benchmarking. Moreover, the code for some of the schedulers is not publicly available, limiting reproducibility and comparative analysis. The following sections present a brief comparison of some schedulers that were tested on the same models.
\subsubsection{\textbf{YOLOv3 for Multi-Frame Execution}}
Table \ref{YOLOv3_non_op} compares the results for the execution of the YOLOv3 model using \ac{Jedi}, \ac{GA}, and \ac{CP-CNN}. The better performance of \ac{Jedi} can be attributed to the horizontal parallelization within the accelerators, emphasizing the importance of proper pipelining and the need for multi-streaming. However, based on Fig. \ref{fig:Jedi_graph}, the increasing throughput across the different models does not cause an increase in power consumption, which is usually not the case. Comparing the performance to \ac{CP-CNN}, \ac{Jedi} utilizes more resources with parallelization, causing a slight difference in energy consumption. Since the idle time is reduced in \ac{CP-CNN}, it offers a more efficient and quicker pipeline. Overall, \ac{CP-CNN} reduces the energy consumed per image by 62\% to 84\% while \ac{Jedi} reduces the energy consumed for the entire set of images up to 55\% \cite{Jedi, CP-CNN}. In the case of \ac{GA}, unlike \ac{Jedi}, it does not consider the use of multi-threading nor multi-streaming, which is the cause of its poor performance compared to \ac{Jedi}, as shown in table \ref{YOLOv3_non_op}. \ac{Jedi} provides a throughput increase of up to 32\% and consumes less energy by up to 55\%, compared to \ac{GA} \cite{Jedi}.
\begin{table}[h]
\caption{Comparison of YOLOv3 execution on NVIDIA Jetson AGX Xavier}
\label{YOLOv3_non_op}
\centering
\begin{threeparttable}
\begin{tabularx}{\columnwidth}{Xccc} % Adjust the width of the table as needed
\toprule
\multirow{2}{*}{Scheduler} & \multicolumn{3}{c}{Performance metrics} \\  
\cmidrule(lr){2-4}
 & Latency (ms) & Throughput (fps) & Energy / Image (mJ) \\ 
\midrule
GPU-only \cite{CP-CNN, Jedi} & 18.10-20.00  & 50-55 & 329-460 \\
\ac{Jedi} \cite{Jedi} & - & 128 & 340 \\
\ac{CP-CNN} \cite{CP-CNN} & 12.96 & - & 306 \\
\ac{GA} \cite{Jedi} & - & 119 & 400 \\
\bottomrule
\end{tabularx}
\begin{tablenotes}
    \item Note: The energy values for \ac{Jedi} and \ac{GA} were calculated approximately by dividing the energy by the number of images used for inference. The results were taken from the mentioned citation.
\end{tablenotes}
\end{threeparttable}
\end{table}
\begin{table}[h]
\caption{Comparison of VGG-19 with ResNet152 execution on NVIDIA Jetson AGX Xavier}
\label{HaX-CoNN_table}
\centering
\begin{threeparttable}
\begin{tabularx}{\columnwidth}{Xccc} % Adjust the width of the table as needed
\toprule
\multirow{2}{*}{Scheduler} & \multicolumn{2}{c}{Performance metrics} \\  
\cmidrule(lr){2-3}
 & Latency (ms) & Throughput (fps) \\ 
\midrule
GPU-only \cite{HaX-CoNN} & 17.05 & 58 \\
Herald \cite{Herald} & 19.73  & 50 \\
\ac{H2H} \cite{H2H} &  16.55 & 60 \\
\ac{HaX-CoNN} \cite{HaX-CoNN} & 13.01 & 77 \\
\bottomrule
\end{tabularx}
\begin{tablenotes}
    \item Note: All values are taken from the reference \cite{HaX-CoNN}.
\end{tablenotes}
\end{threeparttable}
\end{table}

\subsubsection{\textbf{VGG-19 and ResNet152 for Multi-\ac{DNN} Execution}}
\ac{HaX-CoNN} was able to outperform its heuristic predecessors, \ac{H2H} \cite{H2H} and Herald \cite{Herald}, as evidenced by table \ref{HaX-CoNN_table}. The difference in performance between \ac{HaX-CoNN}, \ac{H2H}, and Herald arises from the fact that Herald and \ac{H2H} do not consider memory contention as part of identifying the ideal partitioning point. Although \ac{H2H} considers transition costs, this is insufficient as the latency is impacted by the overhead incurred from the concurrent execution of \ac{DNN}s.
\vspace{-0.8em}
\section{Discussion and Future Prospects}
% Short summary of the paper
Overall, the general tendency is to develop schedulers that use optimization techniques with a single transition point between the \ac{DLA} and GPU. Many determine the ideal schedule using the latency, with some including energy as part of their criteria. As numerous applications require more than a single \ac{DNN} for execution, there is more emphasis in developing schedulers that can handle multiple \ac{DNN} instances. Among schedulers, \ac{HaX-CoNN} is remarkably versatile, accommodating diverse scheduling scenarios, and Map-and-Conquer is adaptable in terms of model architecture. Tables \ref{Summary} and \ref{Summary_costs} summarize all the schedulers reviewed in this paper. The subsequent sections explore challenges that need to be addressed in future works. 
\vspace{-0.8em}
\subsection{Exploring \ac{PVA} and \ac{VIC}}
It is essential to offload tasks or layers, where possible, to the unused \ac{PVA} and \ac{VIC}. In most literature, there is no mention of these devices being employed. One paper explored the relationship between energy and memory contention in CPU, GPU, and \ac{PVA} \cite{MEPHESTO}, which could be utilized in schedulers to ensure an efficient use of \ac{PVA}. A possible pipeline could involve \ac{VIC} for pre-processing tasks like rescaling and image conversion, and \ac{PVA} for initial convolutional layers or filter operations. However, NVIDIA's frameworks (DeepStream and \ac{VPI}) limit the use of the \ac{PVA} to tracking, filtering, and other pre-processing tasks. Using TensorRT, a \ac{DNN} model's layers can only be assigned to the \ac{DLA} or GPU \cite{deepstream, tensorrt}.
\vspace{-0.8em}
\subsection{Power Estimation}
Additionally, there is a need to consider the source of the power and energy statistics. It is unclear where some of the schedulers (\cite{Map_&_Conquer, AxoNN} ) get the power and energy metrics. Typically, the Tegrastats tool is used to measure power consumption \cite{VIC_and_power}. Despite this, the tool often underestimates the actual power consumption \cite{power-estimation-external}, which can impair energy-sensitive systems.
\vspace{-0.8em}
\subsection{Transformer Models}
Transformer-based schedulers are being developed as the popularity of LLMs continues to rise. An example of this is the Map-and-Conquer scheme. Although we can picture the concurrent execution with DNN models, with transformer models, this might be harder to achieve with the increasing memory and computation requirements. Additionally, it is unclear if there is a performance improvement if the model is scheduled vertically rather than horizontally. Further experimentation is required to assess the performance of transformer models in vertical scheduling.
\vspace{-0.8em}
\subsection{\ac{RL} for Scheduling on NVIDIA Edge Devices}
\begin{table*}[!htbp]
\centering
\renewcommand{\arraystretch}{1.3}
\caption{Summary of the schedulers reviewed in this paper}
\label{Summary}
\begin{tabularx}{\textwidth}{@{} l c c c m{1.9cm} >{\centering\arraybackslash}m{1.5cm} m{3cm} m{3cm} @{}}
\toprule
\textbf{Reference} & \textbf{Scheduler} & \textbf{Inter/Intra/Both} & \textbf{S/D} & \textbf{Hardware} & \textbf{Execution} & \textbf{Methodology} & \textbf{Algorithm Aim} \\ 
\midrule

\cite{LP}, 2019 & \ac{LP} & Both & S & \raggedright \arraybackslash GPU, \ac{DLA}, CPU (execute) & Multi-\ac{DNN} & \raggedright \arraybackslash Linear programming model & \raggedright\arraybackslash Minimize overall latency \\ 

\cite{Kang-2020}, 2020 & \ac{GA} & Both & S & \raggedright \arraybackslash GPU, \ac{DLA}, CPU (process) & Multi-frame/\ac{DNN}, Vertical & \raggedright \arraybackslash Parallelization within the accelerator and pipelining between the accelerator using a \ac{GA} & \raggedright\arraybackslash Highest throughput with low energy consumption \\ 

\cite{Herald}, 2021 & Herald & Inter & S & \raggedright\arraybackslash GPU, \ac{DLA} & Multi-\ac{DNN}, Vertical & \raggedright\arraybackslash Profiling followed by design-space exploration & \raggedright\arraybackslash Reduce latency \\ 

\cite{H2H}, 2022 & H2H & Inter & S & \raggedright\arraybackslash GPU, \ac{DLA} & Multi-\ac{DNN}, Vertical & \raggedright\arraybackslash Profiling followed by a set of algorithms to tackle transition cost and latency & \raggedright\arraybackslash Reduce latency and optimize weight locality \\ 

\cite{H2H}, 2022 & Dynamic H2H & Inter & D & \raggedright\arraybackslash GPU, \ac{DLA} & Multi-\ac{DNN}, Vertical & \raggedright\arraybackslash Profiling followed by a set of algorithms to tackle transition cost and latency & \raggedright\arraybackslash Reduce latency and optimize weight locality \\ 

\cite{Jedi}, 2022 & \ac{Jedi} & Both & S & \raggedright \arraybackslash GPU, \ac{DLA}, CPU (process) & Multi-frame, Vertical & \raggedright \arraybackslash Parallelization within the accelerator and pipelining between the accelerator & \raggedright\arraybackslash Highest throughput with satisfactory utilization \\ 

\cite{AxoNN}, 2022 & \ac{AxoNN} & Inter & S & \raggedright\arraybackslash GPU, \ac{DLA} & Single \ac{DNN}, Vertical & \raggedright\arraybackslash Empirical modeling with Z3 SMT solver & \raggedright\arraybackslash Minimize latency while ensuring energy consumption is within a threshold \\ 

\cite{CP-CNN}, 2023 & \ac{CP-CNN} & Inter & S & \raggedright\arraybackslash GPU, \ac{DLA}, CPU (process) & Multi-frame, Vertical & \raggedright\arraybackslash Parallel processing of sequential streaming data & \raggedright\arraybackslash Time to execute part on \ac{DLA} = Time to execute part on GPU \\

\cite{Map_&_Conquer}, 2023 & Map-and-Conquer & Inter & S & \raggedright\arraybackslash GPU, \ac{DLA}, CPU (execute) & Single \ac{DNN}, Horizontal & \raggedright\arraybackslash Parallel processing of inference stages using evolutionary algorithm & \raggedright\arraybackslash Minimizes latency and energy consumption based on thresholds\\

\cite{MaGNAS}, 2023 & \ac{MaGNAS} & Inter & S & \raggedright\arraybackslash GPU, \ac{DLA} & Single \ac{GNN}, Vertical & \raggedright\arraybackslash Two-step evolutionary algorithm & \raggedright\arraybackslash Minimizes latency and energy consumption based on thresholds\\

\cite{HaX-CoNN}, 2024 & \ac{HaX-CoNN} & Inter & S & \raggedright\arraybackslash GPU, \ac{DLA} & Multi-\ac{DNN}, Vertical & \raggedright\arraybackslash Profiling followed by Z3 SAT solver & \raggedright\arraybackslash Maximizes utilization and reduces maximum latency \\ 

\cite{HaX-CoNN}, 2024 & \ac{D-HaX-CoNN} & Inter & D & \raggedright\arraybackslash GPU, \ac{DLA}, CPU (schedule) & Multi-\ac{DNN}, Vertical & \raggedright\arraybackslash Profiling followed by Z3 SAT solver & \raggedright\arraybackslash Maximizes utilization and reduces maximum latency \\ 

\bottomrule
\end{tabularx}
\begin{tablenotes}
    \item Note: S stands for static scheduling and D stands for dynamic scheduling. 
\end{tablenotes}
\end{table*}
\begin{table}[!htbp]
\centering
\renewcommand{\arraystretch}{1.2}
\caption{Scheduling costs considered for each scheduler}
\label{Summary_costs}
\begin{tabular}{@{} l c c c c c @{}}
\toprule
\textbf{Reference} & \textbf{Scheduler} & \textbf{O/H} & \textbf{TC} & \textbf{E} & \textbf{MC}\\ 
\midrule
\cite{LP}, 2019 & \ac{LP} & O & \ding{55} & \ding{55} & \ding{55} \\ 
\cite{Kang-2020}, 2020 & \ac{GA} & O & \ding{51} & \ding{55} & \ding{55} \\ 
\cite{Herald}, 2021 & Herald & H & \ding{55} & \ding{55} & \ding{55} \\ 
\cite{H2H}, 2022 & H2H & H & \ding{51} & \ding{55} & \ding{55} \\  
\cite{Jedi}, 2022 & \ac{Jedi} & H & \ding{51} & \ding{55} & \ding{55} \\ 
\cite{AxoNN}, 2022 & \ac{AxoNN} & O & \ding{51} & \ding{51} & \textbf{-} \\ 
\cite{CP-CNN}, 2023 & \ac{CP-CNN} & H & \ding{51} & \ding{55} & \ding{55} \\ 
\cite{Map_&_Conquer}, 2023 & Map-and-Conquer & O & \ding{51} & \ding{51} & \ding{55} \\ 
\cite{MaGNAS}, 2023 & \ac{MaGNAS} & O & \ding{51} & \ding{51} & \textbf{-} \\ 
\cite{HaX-CoNN}, 2024 & \ac{HaX-CoNN} & O & \ding{51} & \ding{55} & \ding{51} \\ 
\bottomrule
\end{tabular}
\begin{tablenotes}
    \item Note: O stands for optimization, H for heuristic, TC for transition costs, E for energy consumption, and MC for memory contention. 
\end{tablenotes}
\end{table}
Another avenue to explore is the use of AI for scheduling. For instance, SCHED² \cite{RL-SCHED²} and DL2 \cite{DL2} are \ac{RL} techniques used to determine the best use of resources within a GPU data center \cite{Survey_GPU_Data_Centers}. Moreover, \cite{RL} depicts how \ac{RL} can be used even in heterogeneous edge devices with a CPU-GPU architecture. It simultaneously tackled multiple pre-processing, inference, and training jobs, outperforming existing heuristic methods. Similarly, RESPECT \cite{RESPECT} applies \ac{RL} in edge \ac{TPU} devices to schedule \ac{DNN} models. In this case, the \ac{DNN} is modeled as a directed graph. The agent then learns to generate near-optimal schedules with minimal overhead. Combining the two techniques can be applied to the NVIDIA edge devices, allowing more dynamic scheduling to balance multiple inference workloads on the CPU, GPU, and dedicated accelerators. A priority list can also be set, based on the application. 
\vspace{-0.5em}
\subsection{Real-Time Deployment in Industrial Applications}
Notably, the reviewed schedulers have not been evaluated with live input streams under real-time circumstances. Therefore, future work should prioritize the assessment in real-time conditions with real data. Other critical aspects for industrial implementation are scalability, security, and fault tolerance. In terms of scalability, these schedulers (except for \ac{MaGNAS}) can be implemented for any \ac{DNN} models. For transformer models, as mentioned previously, Map-and-Conquer may be used; however, it may not be possible to implement the scheduling for all transformer model architectures due to limitations of the \ac{DLA}. Another facet of scalability is the ability to implement multiple models. \ac{HaX-CoNN} is an ideal choice as it accounts for memory contention. If more than 3 \ac{DNN} models are running concurrently, opting for the simplistic \ac{LP} scheduler instead would be better. Current schedulers did not consider security and fault tolerance as principal design considerations. They are not immune to faults, as memory crashes can disrupt the pipeline without any way to recover. Developing more kernel-based scheduling techniques may be more suitable to circumvent these issues and operate at a minimum performance. Here, the operating system dynamically schedules the AI models through threads. This would mitigate the risk of pipeline elimination from security threats and interruptions, unlike user scheduling, where the scheduling algorithm is separate from the operating system. Kernel-based scheduling would also improve the system's fault tolerance. 
\vspace{-0.5em}
\subsection{Summary}
\begin{figure}[H]
    \centering
    \includegraphics[width=\linewidth]{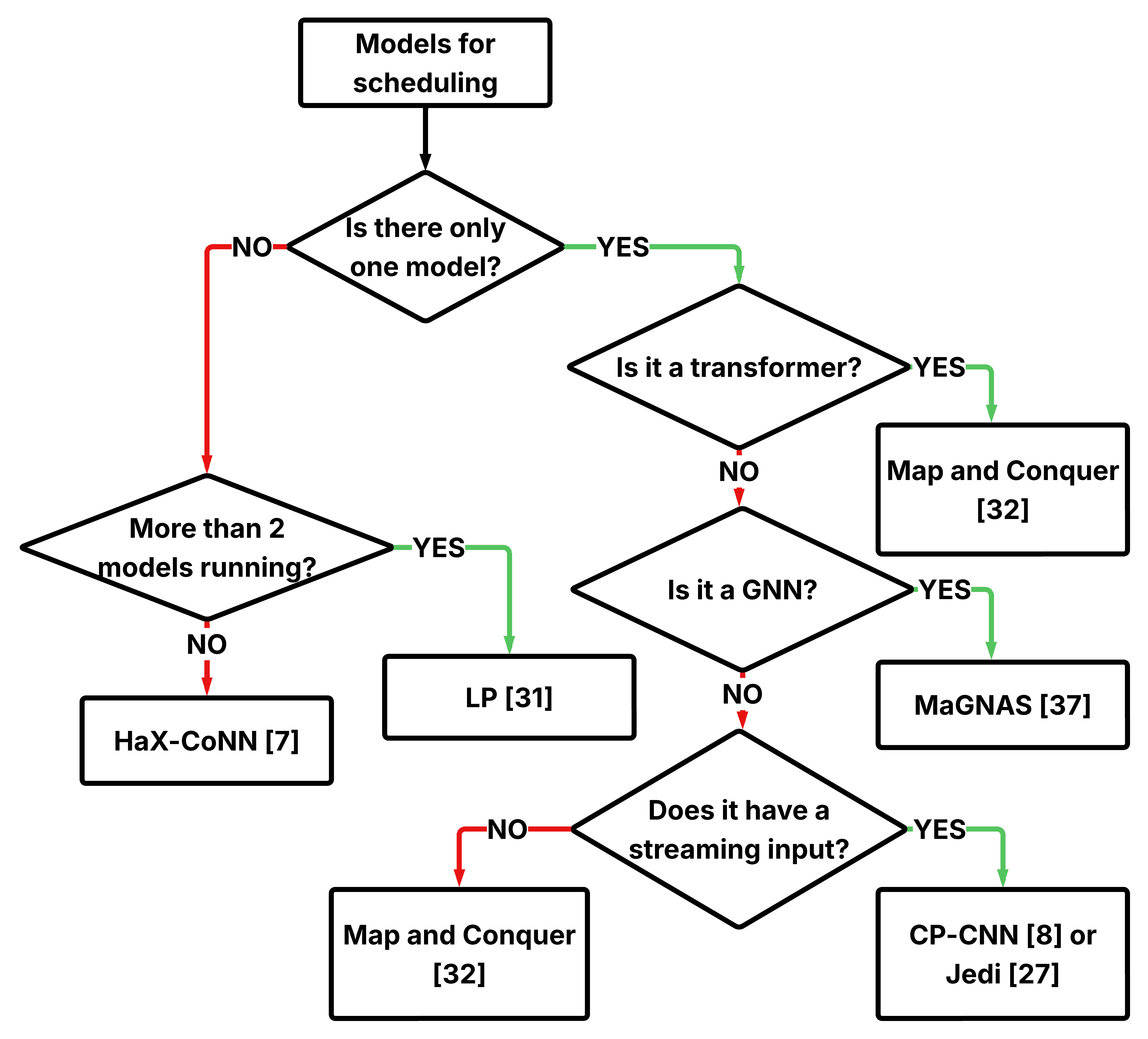}
    \caption{ Flowchart to aid in choosing the ideal scheduler}
    \label{fig:flowchart}
\end{figure}
Overall, the discussion highlights five key insights that emerge from the review and comparative analysis. First, future schedulers should better utilize the underused on-chip accelerators such as the \ac{PVA} and \ac{VIC}, which can offload pre- and post-processing tasks to reduce GPU and \ac{DLA} workloads. Second, more accurate methodologies for power estimation are needed, as tools such as Tegrastats tend to underestimate real energy consumption, limiting the reliability of energy-aware optimization. Third, the integration of transformer models introduces new scheduling challenges due to their high memory and computation requirements, calling for further exploration of both vertical and horizontal scheduling strategies. Fourth, reinforcement learning presents a promising avenue for developing adaptive and dynamic scheduling schemes that can handle heterogeneous workloads efficiently across CPU, GPU, and AI accelerators. Finally, real-time deployment remains a critical gap in the current literature—few schedulers have been tested with live input streams or evaluated for scalability, security, and fault tolerance, all of which are essential for industrial implementation. Tables~\ref{Summary} and~\ref{Summary_costs} summarize the general trends, showing that most schedulers primarily optimize for latency with limited treatment of transition costs, energy, and memory contention.

Ultimately, the choice of the scheduler must align with the requirements of the target application, as illustrated by Fig. \ref{fig:flowchart}. As discussed, in the case of multiple \ac{DNN}s, it is better to opt for \ac{HaX-CoNN} where possible, or resort to \ac{LP}. With a single \ac{DNN} carrying out real-time processing, \ac{CP-CNN} and \ac{Jedi} are ideal choices since they allow for multi-frame execution. If energy consumption is a concern, Map-and-Conquer is a better option. Currently, Map-and-Conquer is the only choice for transformer models, while for \ac{GNN} models, \ac{MaGNAS} is the sole solution for edge devices. These schedulers remain applicable in these devices, even if the architecture changes with new processing units or an updated \ac{DLA}, as long as the device uses a unified, shared memory system.
\vspace{-0.2em}
\section{Conclusion}
This paper systematically compares the different schedulers developed for the more recent NVIDIA Jetson edge devices that include a \ac{DLA} in its hardware. Some of the research challenges that need to be considered are the further development of the TensorRT layer assignment tools to allocate layers to the CPU and PVA. This can open more possibilities for scheduling and may even be more efficient than just using the GPU and DLA. For successful implementation, this would require studies into the memory contention and latency-energy trade-offs, similar to \cite{MEPHESTO} and \cite{PCCS}, while ensuring that the source of the power statistics is accurate. Exploring layer substitutions to target DLA-incompatible layers within the model would also be beneficial. To improve the security and fault tolerance of the system, it might be more advantageous to combine an RL-based scheduler with the operating system to create a more resilient system. The focus of future scheduling techniques should be to balance multiple AI models as an increasing number of robotics applications demand multi-modal systems. Furthermore, the schedulers need to be tested extensively in real-time applications.
\vspace{-0.4em}
\ifCLASSOPTIONcaptionsoff
  \newpage
\fi

% trigger a \newpage just before the given reference
% number - used to balance the columns on the last page
% adjust value as needed - may need to be readjusted if
% the document is modified later
%\IEEEtriggeratref{8}
% The "triggered" command can be changed if desired:
%\IEEEtriggercmd{\enlargethispage{-5in}}

% references section

% can use a bibliography generated by BibTeX as a .bbl file
% BibTeX documentation can be easily obtained at:
% http://mirror.ctan.org/biblio/bibtex/contrib/doc/
% The IEEEtran BibTeX style support page is at:
% http://www.michaelshell.org/tex/ieeetran/bibtex/
%\bibliographystyle{IEEEtran}
% argument is your BibTeX string definitions and bibliography database(s)
%\bibliography{IEEEabrv,../bib/paper}
%
% <OR> manually copy in the resultant .bbl file
% set second argument of \begin to the number of references
% (used to reserve space for the reference number labels box)
\bibliographystyle{IEEEtran}
\bibliography{REF_TII-25-4607-R1}

% biography section
% 
% If you have an EPS/PDF photo (graphicx package needed) extra braces are
% needed around the contents of the optional argument to biography to prevent
% the LaTeX parser from getting confused when it sees the complicated
% \includegraphics command within an optional argument. (You could create
% your own custom macro containing the \includegraphics command to make things
% simpler here.)
%\begin{IEEEbiography}[{\includegraphics[width=1in,height=1.25in,clip,keepaspectratio]{mshell}}]{Michael Shell}
% or if you just want to reserve a space for a photo:

%\begin{IEEEbiography}{Michael Shell}
%Biography text here.
%\end{IEEEbiography}

% if you will not have a photo at all:
%\begin{IEEEbiographynophoto}{John Doe}
%Biography text here.
%\end{IEEEbiographynophoto}

% insert where needed to balance the two columns on the last page with
% biographies
%\newpage

%\begin{IEEEbiographynophoto}{Jane Doe}
%Biography text here.
%\end{IEEEbiographynophoto}

% You can push biographies down or up by placing
% a \vfill before or after them. The appropriate
% use of \vfill depends on what kind of text is
% on the last page and whether or not the columns
% are being equalized.

%\vfill

% Can be used to pull up biographies so that the bottom of the last one
% is flush with the other column.
%\enlargethispage{-5in}

% that's all folks
\end{document}